# Ultrafast ion sieving in two dimensional graphene oxide membranes

Junfan Liu[1,3,†], Zonglin Gu[1,†,*], Mengru Duan[1], Pei Li[2], Lu Li[3], Jianjun Jiang[1], Rujie Yang[3], Junlang Chen[3], Zhikun Wang[3], Liang Zhao[1], Yusong Tu[1,*], Liang Chen[2,*]

[1]*College of Physical Science and Technology & Microelectronics Industry Research Institute, Yangzhou University, Jiangsu, 225009, China*
[2]*School of Physical Science and Technology, Ningbo University, Ningbo 315211, China*
[3] *Department of Optical Engineering, Zhejiang Prov Key Lab Carbon Cycling Forest Ecosy, College of Environmental and Resource Sciences, Zhejiang A&F University, Hangzhou 311300, China.*

**Ultrahigh water permeance, together with a high rejection rate through nanofiltration and separation membranes[1,2], is crucial but still challenging for multivalent ion sieving in water treatment processes of desalination, separation, and purification[3,4]. To date, no theory or equation has ever been quantitatively clarified the mechanism of water permeance in two-dimensional (2D) membranes, despite intensive and prolonged searches. Here, we established a new general equation of permeation through 2D membranes, and experimentally achieved unprecedented advances in water permeance one to two orders of magnitude higher than state-of-the-art membranes while simultaneously maintaining high ion rejection rates for multivalent metal ions, by staking nano-sized reduced graphene oxide (nano-rGO) flakes into nanofiltration membranes. The equation is simply based on a fundamental steady-state flow assumption and provides an essential description of water permeance through 2D membranes, demonstrating that the ultrahigh water permeance is attributed to the high effective channel area and shortened channel length elicited from the nano-sized-flake stacking effects in nano-rGO membranes, consistent with our theoretical simulations and previous experiments. These results pave the way for fabrication of advanced 2D nanofiltration membranes to realize a breakthrough in water permeance with exceptional ion sieving performance.**

Removal of toxic heavy metal ions is important for water separation, recovery, and purification from seawater, inland brackish water, and various waste water[3,5] in order for the water supply to cope with the growing freshwater crisis[6,7]. Nanofiltration membranes offer significant advantages of high energy efficiency and low cost effectiveness and have attracted increasingly extensive interest for ion removal treatment processes[4,8], particularly with the use of two-dimensional (2D) materials[9,10], such as graphene oxide (GO)-based membranes with precise ionic sieving[1,11,12]. Currently, the development of membrane technologies with substantially improved water permeance for effective removal of ions is considered one of the most important objectives in nanofiltration separation applications[6,9].

2D nanofiltration membranes have great potential for ultrafast water permeance with effective ion rejection due to their atom-thick features[1,4,13]. Recently, numerous efforts have been made to employ diverse 2D materials, such as the GO[14,15], $MoS_2$[16,17], and MXene[18,19] families, to fabricate 2D membranes and explore their relevant characteristics to achieve ultrafast water permeance, including tuning porous microstructures for water transport[20,21], adjusting interfacial hydrophobicity for low water friction[22,23], and optimizing membrane parameters (e.g., thickness and material lateral sizes)[15,24]. Despite these significant advances, water permeance through state-of-the-art nanofiltration membranes tends to be on the order of a dozen or dozens L m$^{-2}$ h$^{-1}$ bar$^{-1}$ (LMH bar$^{-1}$)[14,25] for multivalent metal ion removal, and the highest permeance reported in the literature was just ~164.7 LMH bar$^{-1}$, even with an ion rejection rate of 85.2%[22]. Although very promising, further improvement in permeance appears to be hindered, partially due to the lack of essential understanding of water permeation through 2D membranes. Currently, no theory or equation is available to quantitatively clarify the relationship between water permeance and relevant membrane parameters. Hagen–Poiseuille equation is usually used to evaluate flow permeation through nanofiltration membranes (including 2D membranes), but its permeance predictions present a huge departure from experiments[26-29] (i.e., four to six orders of magnitude smaller than experimental values). Therefore, not only does it remain a great challenge to substantially upgrade water permeance in nanofiltration membranes, but a comprehensive theory or framework for guiding the development of filtration and separation membranes is urgently expected, particularly in the arising field of 2D membranes.

Here, we combined experimental and theoretical approaches to demonstrate ultrafast multivalent ion sieving in nano-rGO membranes stacked with nano-sized rGO flakes. The water permeance reached up to 1112 LMH bar$^{-1}$ with a correspondingly high rejection rate of 91.4%, which is one to two orders of magnitude higher than the permeance of state-of-the-art membranes. Herein, a new general equation was established to provide an essential description and accurate evaluation of water permeance in 2D membranes that is consistent with our theoretical simulations and filtration experiments and validated by a series of previous experiments.

We prepared a nano-rGO suspension from a GO suspension with micron-sized GO flakes via a hydrothermal method. The nano-rGO membranes were prepared by stacking the obtained nano-rGO suspension on a mixed cellulose ester (MCE) substrate with a pore size of 0.22 μm (see Fig. 1a and Supplementary Information section PS1). The nano-rGO flakes had an average lateral size of ~100 nm with a size distribution range of ~50 nm to 250 nm and thickness of ~1.0 nm (AFM measurements, Fig. 1b). The lateral size of the nano-rGO flakes was reduced by a factor of approximately 10 compared to the original GO flakes. The oxygen content of the nano-rGO membrane decreased significantly from 35.5% to 26.5% (indicated by XPS measurements in Fig. 1d and Table S1), with the corresponding water contact angle increasing from 39.7º to 50.5º compared to the GO membrane (Fig. S1b), indicating that the nano-rGO hydrophobicity was significantly enhanced. In previous work, it was difficult to form a stable membrane directly with nano-sized GO flakes because the hydrophilic GO flakes with lateral sizes less than the substrate pore sizes made them highly susceptible to swelling and disintegration[10,24]. The enhanced hydrophobicity can lead to the formation and stability of nano-rGO membranes stacked on the MCE substrate. Scanning electron microscopy (SEM) images (Fig. 1c) showed that the stacked nano-rGO membrane is continuous and free of macro pores or defects, which is critical for a highly efficient separation process[30].

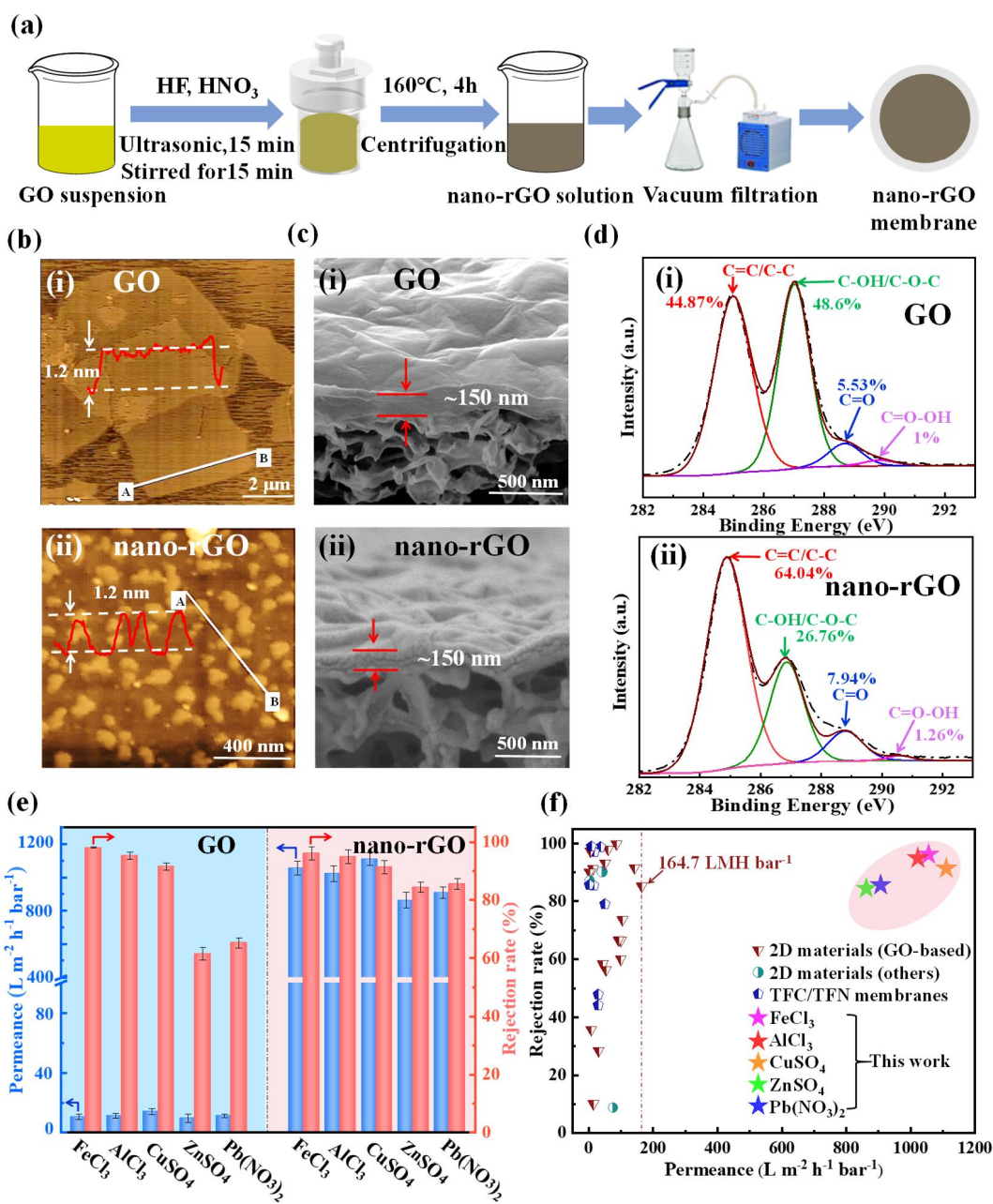

**Figure 1. Multivalent ion sieving in reduced graphene oxide (rGO) membrane stacked with nano-sized rGO (nano-rGO) flakes. a,** Schematic of the fabrication of the nano-rGO membrane. **b**, Atomic force microscopy (AFM) images with corresponding height profiles taken along the marked white line. (i) Graphene oxide (GO) and (ii) nano-rGO flakes. **c,** Cross-sectional scanning electron microscope (SEM) images of GO (i) and nano-rGO membrane (ii). **d**, X-ray photoelectron spectra (XPS) of C1s for GO (i) and nano-rGO (ii). **e**, Water permeance and rejection rates of GO and nano-rGO membranes for 50 mg/L $FeCl_3$, $AlCl_3$, $CuSO_4$, $ZnSO_4$, and $Pb(NO_3)_2$. Error bars indicate the standard deviation determined from three different samples. **f**, Comparison of the water permeance and salt rejection performance of the nano-rGO membrane with state-of-the-art membranes reported in the literature for multivalent ion removal (brown inverted triangle, GO-based membranes; cyan circles, other two-dimensional material membranes; dark blue pentagon, TFN/TFC membranes; details in Supplementary Information section PS3).

Nanofiltration experiments demonstrated excellent performance of ultrafast ion sieving in the nano-rGO membrane. As shown in Fig. 1e, the water permeance through nano-rGO membranes reached up to 1057 ± 42 LMH bar$^{-1}$, 1023 ± 47 LMH bar$^{-1}$, 1112 ± 38 LMH bar$^{-1}$, 863 ± 47 LMH bar$^{-1}$, and 908 ± 35 LMH bar$^{-1}$ for $FeCl_3$, $AlCl_3$, $CuSO_4$, $ZnSO_4$, and $Pb(NO_3)_2$, with corresponding rejection rates of 96.2 ± 2.3%, 94.9 ± 2.4%, 91.4 ± 2.2%, 84.4 ± 1.8%, and 85.6 ± 1.8%, respectively. In contrast, the permeance values through GO membranes were close to 10 LMH bar$^{-1}$ with moderate rejection rates. Thus, the nano-rGO membrane presents a significant increase in water permeance over GO membranes by two orders of magnitude. Notably, the highest water permeance recently reported for state-of-the-art nanofiltration membranes was only 164.7 LMH bar$^{-1}$, even with a rejection rate of 85.2% for multivalent metal ions[22] (Fig. 1f and Table S2). In comparison, the water permeance of the nano-rGO membrane in this study was one to two orders of magnitude higher than the permeance of state-of-the-art nanofiltration membranes, including 2D membranes (GO-based, $MoS_2$, and MXene), thin-film composite (TFC), and thin-film nanocomposite (TFN) membranes.

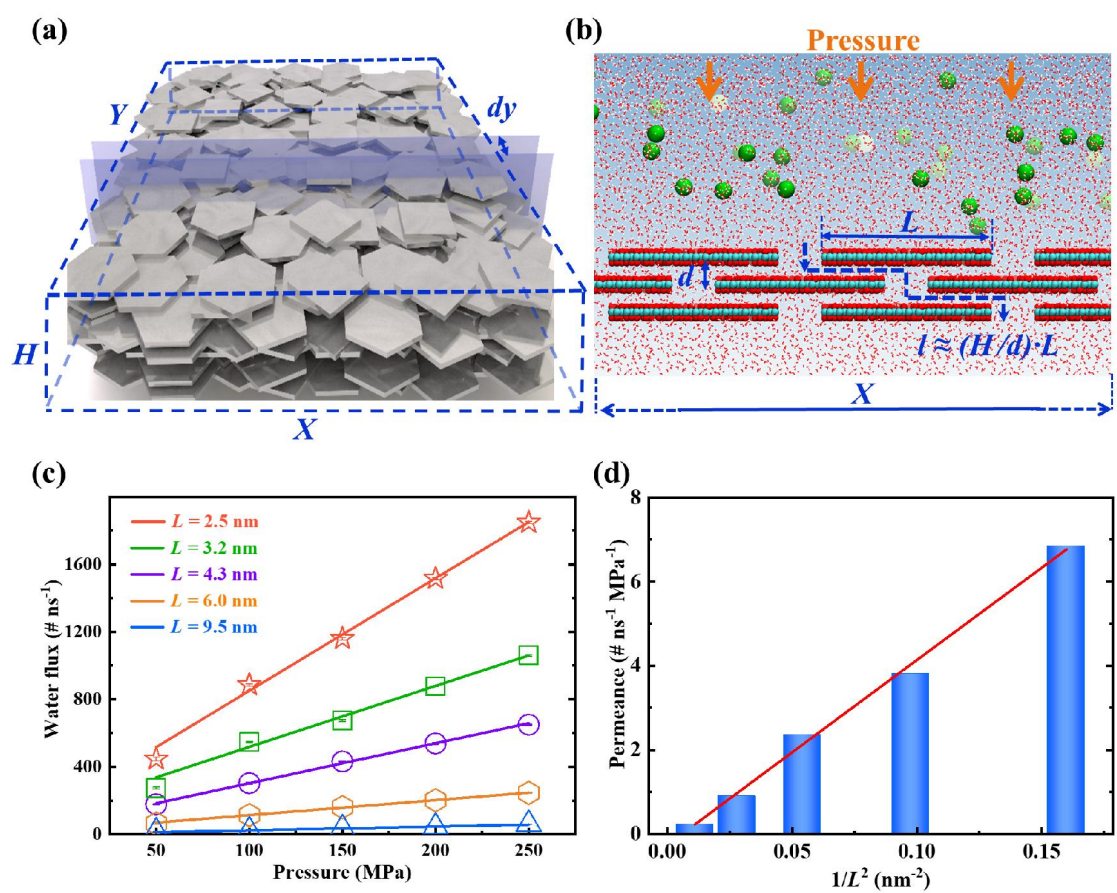

**Figure 2. Simulations of the effects of flake lateral sizes on water flux through nano-rGO membrane. a,** Schematic showing the membrane randomly stacked with nano-rGO flakes and one cross-section with an infinitesimal width. $X$ and $Y$ indicate two-dimensional sizes of the membrane plane: $H$, the membrane thickness, and $dy$, the cross-section infinitesimal width along $Y$, respectively. **b,** Physical model of the membrane constructed by elongating the width of the cross-section along

the *y* direction, statistically equivalent to the randomly stacked membrane. The lateral sizes of the flakes are set the same as the average *L*, and the interlayer spacings are set uniformly as the average *d*. The dashed lines indicate one water channel with its effective length of *l*. The carbon and oxygen atoms of the nano-rGO flakes were shown in cyan and red, water molecules in red, and multivalent ions in green. External pressures were exerted on the feed side. **c,** Pressure dependence of water flux through the membranes (water number per nanosecond). **d,** Linear relationship between the water permeance (water flux per unit pressure) and the reciprocal of the square of the lateral size.

Molecular dynamics (MD) simulations demonstrated that the ultrafast water permeance through the membranes is proportional to the reciprocal of the square of the lateral size of the flakes. Fig. 2a shows a schematic of the membrane stacked by random self-assembly of nano-rGO flakes, with dimensional sizes (*X* and *Y*) of the membrane plane and the membrane thickness (*H*). We selected a cross-section of the membrane (*X*×*H*) with its infinitesimal width *dy*. Accordingly, as shown in Fig. 2b, by elongating its width along the *y* direction, we constructed a physical model that has water permeance statistically equivalent to the permeance of the randomly stacked membranes (see the deduction of the equation (1) below). The lateral size of the nano-rGO flakes is considered to be the same as the average size (*L*), and the interlayer spacings are set uniformly to the average (*d*). We applied this physical model membrane (with the different lateral sizes of flakes, $L = 2.5 \sim 9.5$ nm, and external pressure, $P = 50 \sim 250$ MPa) to perform MD simulations (details in PS4). Fig. 2c shows that the flux increased linearly with increasing pressure for the membrane with a given flake size, indicating the steady-state water flow through the membranes, whereas the flux increased rapidly with decreasing flake size. Furthermore, as shown in Fig. 2d, we calculated the flux per unit pressure as the permeance and found a clear linear relationship between the water permeance and the reciprocal square of the lateral size ($1/L^2$). In our experiments, nano-rGO and GO membranes had the same thickness and the lateral sizes of their flakes were approximately 100 nm and 1 μm, respectively. According to this relationship, we estimated the permeance and found the permeance ratio between nano-rGO and GO membranes is approximately 100. Remarkably, this theoretical fold of two orders of magnitude is consistent with our filtration experiment results (Fig. 1e).

Theoretically, we established a new general equation for evaluating the water permeance through 2D membranes. As a general case, we considered a steady-state flow through the membranes and assumed an average flow velocity proportional to the pressure gradient exerted on the membranes: $\bar{v} = v \cdot P/l$, where *P* is the pressure, *l* is the effective channel length, and *v* is a coefficient representing an average flow velocity per unit pressure gradient. Thus, the flux per unit membrane area and unit pressure, i.e., permeance, is written as $J_w = v/l \cdot S$, where *S* is the effective channel area per unit membrane area. Careful examination of the physical model of 2D membranes in Fig. 2b showed that, for the selected cross-section, the number of water channels can be estimated as $n = X/L$, and the effective area for one water channel, $ds = d \cdot dy$; considering that each selected cross-section of the membrane statistically has the same *n* and all of the channels are aligned and elongated along the *y* direction, the sum of all effective channel areas can be deduced as: $S_{sum} = \int n \cdot ds = XY \cdot d/L$. Thus, $S = d/L$, per unit area of the whole membrane. Moreover, the effective channel length is $l = (H/d) \cdot L = H \cdot L/d$. Therefore, the permeance equation for 2D membranes is as in equation (1):

$$J_w = v/l \cdot S = v \cdot d^2/(H \cdot L^2) \qquad (1)$$

The above deduction process indicates that, as the lateral size of the flake becomes small, the small-flake stacking can elicit advantageous effects (i.e., increasing channel number ($n$) and the subsequent effective channel area, as well as shortening the effective channel length), which significantly increases the water permeance. Quantitatively, the permeance is proportional to the reciprocal of the square of the lateral size of the flake ($1/L^2$) for a 2D membrane with given $H$ and $d$, which is consistent with the results of the MD simulations and filtration experiments above. Moreover, the water permeance $J_w$ has an inverse relationship with the membrane thickness $H$ for a membrane with given $L$ and $d$, which was also confirmed in both MD simulations and filtration experiments (Fig. S4-S5 and PS5-PS6).

We note that Hagen–Poiseuille equation is usually employed to evaluate the flow permeation through 2D membranes but fails to accurately estimate the permeance, with results four to six orders of magnitude smaller than the values detected in experiments[26-29]. This huge departure from experiments were thought to be mainly attributed to large slip flow through 2D nanocapillaries[9,12], while the slip flow theory remains a controversial topic due to a lack of sufficient direct proof[9,28,30]. Equation (1) above is based on the steady-state flow assumption in 2D membranes, overlooking some specific velocity profiles and introducing an average velocity per unit pressure gradient as a coefficient, $v$. Statistically, this coefficient is representative of flow permeability through membranes and is generally dominated by the characteristics of the fluid (e.g., viscosity) and membrane flake materials (e.g., surface chemistry, wetting, and complicated boundary), as well as their interfacial interactions. Table 1 shows the experimental determination of permeability coefficient $v$ calculated from Equation (1) in terms of the experimental permeance and corresponding parameters of currently existing 2D membranes, including the GO, $MoS_2$, and MXene membrane families. Interestingly, permeability coefficient $v$ exhibits excellent consistency for each type of 2D membrane, indicating that $v$ is an intrinsic characteristic of 2D membranes. Upon the permeability coefficient $v$, Equation (1) allows us to quantitatively describe water permeation through 2D membranes and further enhance the permeability by ingenious optimization. Importantly, the equation is deduced directly from both fundamental physical concepts of effective channel area and effective channel length, providing essential and crucial insights into the ultrafast sieving permeance of 2D membranes in nanofiltration applications.

**Table 1. Experimental determination of permeability coefficient *v* through two-dimension membranes for ionic sieving in the literature.**

| Membrane type | Average lateral size $L$ (nm) | Interlayer spacing $d$ (nm) | Membrane thickness $H$ (nm) | Permeance $J_w$ (LMH bar$^{-1}$) | **Permeability coeff.*** $v$ (m$^2$ h$^{-1}$ bar$^{-1}$) | Ref. |
|---|---|---|---|---|---|---|
| GO | 1000 | 1.2 | 150 | ~10 | **1.0** | This work |
| GO | 1000 | 1.2 | 150 | ~10 | **1.0** | [22] |
| GO | 1000 | 1.2 | 150 | ~12 | **1.3** | [14] |
| GO-PEI | 1000 | 1.2 | 200 | ~8 | **1.1** | [14] |
| GO-PAA | 1000 | 1.2 | 200 | ~8 | **1.1** | [14] |
| GO | 10000 | 1.14 | 50 | ~0.5 | **1.9** | [15] |
| rGO | 1000 | 1.15 | 330 | ~11 | **2.7** | [31] |
| nano-rGO | 100 | 1.1 | 150 | ~1112 | **1.6** | This work |
| GO-FLG | 1500 | 1.4 | 30 | ~12.6 | **0.4** | [21] |
| GO-COOH | 1000 | 1.41 | 50 | ~20 | **0.5** | [32] |
| MoS$_2$ | 304 | 1.2 | 500 | ~50 | **1.6** | [17] |
| MoS$_2$-EtOH | 300 | 1.1 | 500 | 13.4 | **0.5** | [16] |
| MoS$_2$-Methyl | 300 | 1.08 | 500 | 15.9 | **0.6** | [16] |
| PEI-MoS$_2$ | 350 | 0.91 | 1800 | ~4.6 | **1.2** | [33] |
| MXene | 1000 | 1.47 | 50 | ~30 | **0.7** | [34] |
| SC-MXene | 1000 | 1.47 | 80 | ~9 | **0.3** | [34] |
| MXene | 700 | 0.83 | 150 | ~4.7 | **0.5** | [35] |
| MXene | 1000 | 1.6 | 80 | ~13 | **0.4** | [18] |
| Al$^{3+}$-MXene | 1000 | 1.52 | 1100 | ~2.8 | **1.3** | [19] |

GO-PEI, GO membrane with surface coating of polyethylene imine; GO-PAA, GO membrane with surface coating of polyacrylic acid; rGO, reduced GO membrane; GO-FLG, mixed GO and few-layered graphene membrane; GO-COOH, carboxyl functionalized GO membrane; MoS$_2$-EtOH, ethyl-2-ol- functionalized MoS$_2$ membrane; MoS$_2$-Methyl, methyl functionalized MoS$_2$ membrane; PEI-MoS$_2$, MoS$_2$ and polyethylene imine composite membrane; SC-MXene, surface charged MXene membrane; Al$^{3+}$-MXene, Al$^{3+}$ intercalated MXene membrane.

*$v$ is calculated from Equation (1) with realistic parameters and experimental permeance.

We also demonstrated the excellent robustness and durable stability of our nano-rGO membranes. Fig. 3a presents the long-term operation schematic with a crossflow-like method. The salt solution was transported to the feed by a peristaltic pump to maintain the solution volume in the feed, forming a cyclic flow in the whole device, and a stirrer was used in the salt solution in feed to achieve uniform ion concentration. Experiments under a pressure of 1 bar showed long-term stability of the nano-rGO membrane for 240 minutes (Fig. 3b). The water permeance decreased slightly from 1270 LMH bar$^{-1}$ within the first 100 minutes and then held steady at 952 LMH bar$^{-1}$ after reaching equilibrium. The corresponding rejection rates were eventually stable around 92.9%. These results clearly show the outstanding stability of our nano-rGO membranes with excellent filtration performance.

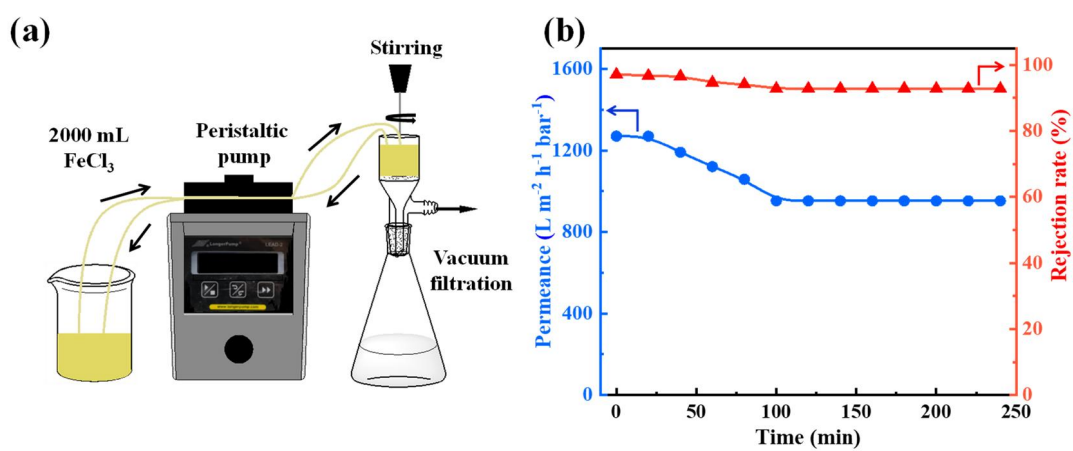

**Figure 3. Long-term stability of nano-rGO membranes conducted under crossflow-like conditions. a,** Nanofiltration setup with crossflow-like conditions. **b,** Water permeance and salt rejection rates in nano-rGO membranes for $FeCl_3$ sieving over a 240 min operating period under a transmembrane pressure of 1 bar.

In summary, using both experimental and theoretical approaches, we have achieved ultrafast multivalent ion sieving in a nano-rGO membrane stacked with nano-sized rGO flakes. Remarkably, the nano-rGO membrane exhibited a significant advance in water permeance, one to two orders of magnitude higher than state-of-the-art membranes, while maintaining high ion rejection rates. This significant advance in nanofiltration performance was first predicted theoretically by Equation (1), which we established here, and then achieved experimentally in our nano-rGO membranes. The ultrahigh water permeance is attributed to both the high effective channel area and the shortened channel length elicited from nano-sized-flake stacking effects in nano-rGO membranes.

Notably, this new general equation is simply based on a steady-state flow assumption for water permeance through 2D membranes. Different from Hagen–Poiseuille equation, we introduced an average velocity per unit pressure gradient as permeability coefficient $v$ and quantitatively clarified the relationship between water permeance and relevant membrane parameters. We noted that the interlayer spacing is usually limited to adjust to and restricted strictly for the ion rejection requirement in 2D membranes. In this scenario, the clear experimental determination of the permeability coefficient ($v$) in Table 1 indicates that Equation (1) offers a sufficient and essential description of water permeance through 2D membranes for ion sieving. Here, the permeability coefficient in 2D membranes is indicated to be determinative of the intrinsic characteristics of membranes that usually result from 2D flake materials and their fabrication; the decrease in lateral size of the flakes and the membrane thickness is necessary to increase the number of effective water channels per unit area of membrane and to reduce the effective channel length, upgrading the water permeance of ion sieving in 2D membranes. We believe that the concept of permeability coefficient is also of prime importance in general nanofiltration membranes, which is never put forward before, and the physical essence enables it to be easily extended to various nanofiltration processes. All of the above findings offer fundamental and crucial insights into water permeance through 2D membranes to guide the development of next-generation filtration and separation membranes.

References

1  Joshi, R. K. *et al.* Precise and ultrafast molecular sieving through graphene oxide membranes. *Science* **343**, 752-754, (2014).
2  Park, H. B., Kamcev, J., Robeson, L. M., Elimelech, M. & Freeman, B. D. Maximizing the right stuff: the trade-off between membrane permeability and selectivity. *Science* **356**, eaab0530 (2017).
3  Sholl, D. S. & Lively, R. P. Seven chemical separations to change the world. *Nature* **532**, 435-437, (2016).
4  Shannon, M. A. *et al.* Science and technology for water purification in the coming decades. *Nature* **452**, 301-310, (2008).
5  Li, W. W., Yu, H. Q. & Rittmann, B. E. Chemistry: reuse water pollutants. *Nature* **528**, 29-31, (2015).
6  Werber, J. R., Osuji, C. O. & Elimelech, M. Materials for next-generation desalination and water purification membranes. *Nat. Rev. Mater.* **1**, 16018 (2016).
7  Landsman, M. R. *et al.* Water treatment: are membranes the panacea? *Annu. Rev. Chem. Biomol.* **11**, 559-585 (2020).
8  Elimelech, M. & Phillip, W. A. The future of seawater desalination: energy, technology, and the environment. *Science* **333**, 712-717 (2011).
9  Shen, J., Liu, G., Han, Y. & Jin, W. Artificial channels for confined mass transport at the sub-nanometre scale. *Nat. Rev. Mater.* **6**, 294-312 (2021).
10  Sun, P. Z., Wang, K. L. & Zhu, H. W. Recent developments in graphene-based membranes: structure, mass-transport mechanism and potential applications. *Adv. Mater.* **28**, 2287-2310 (2016).
11  Chen, L. *et al.* Ion sieving in graphene oxide membranes via cationic control of interlayer spacing. *Nature* **550**, 415-418 (2017).
12  Abraham, J. *et al.* Tunable sieving of ions using graphene oxide membranes. *Nat. Nanotechnol.* **12**, 546-550 (2017).
13  Xie, Q. *et al.* Fast water transport in graphene nanofluidic channels. *Nat. Nanotechnol.* **13**, 238-245 (2018).
14  Zhang, M. C. *et al.* Controllable ion transport by surface-charged graphene oxide membrane. *Nat. Commun.* **10**, 1253 (2019).
15  Yang, Q. *et al.* Ultrathin graphene-based membrane with precise molecular sieving and ultrafast solvent permeation. *Nat. Mater.* **16**, 1198-1202 (2017).
16  Ries, L. *et al.* Enhanced sieving from exfoliated MoS2 membranes via covalent functionalization. *Nat. Mater.* **18**, 1112-1117 (2019).
17  Wang, Z. Y. *et al.* Understanding the aqueous stability and filtration capability of MoS2 membranes. *Nano Lett.* **17**, 7289-7298 (2017).
18  Wang, J. *et al.* Ion sieving by a two-dimensional Ti3C2Tx alginate lamellar membrane with stable interlayer spacing. *Nat. Commun.* **11**, 3540 (2020).
19  Ding, L. *et al.* Effective ion sieving with Ti3C2Tx MXene membranes for production of drinking water from seawater. *Nat. Sustain.* **3**, 296-302 (2020).

20 Yang, Y. *et al.* Large-area graphene-nanomesh/carbon-nanotube hybrid membranes for ionic and molecular nanofiltration. *Science* **364**, 1057-1062 (2019).

21 Morelos-Gomez, A. *et al.* Effective NaCl and dye rejection of hybrid graphene oxide/graphene layered membranes. *Nat. Nanotechnol.* **12**, 1083-1088 (2017).

22 Yi, R. B. *et al.* Selective reduction of epoxy groups in graphene oxide membrane for ultrahigh water permeation. *Carbon* **172**, 228-235 (2021).

23 Lu, X. L. & Elimelech, M. Fabrication of desalination membranes by interfacial polymerization: history, current efforts, and future directions. *Chem. Soc. Rev.* **50**, 6290-6307 (2021).

24 Nie, L. *et al.* Realizing small-flake graphene oxide membranes for ultrafast size-dependent organic solvent nanofiltration. *Sci. Adv.* **6**, eaaz9184 (2020).

25 Dai, F. F. *et al.* Ultrahigh water permeation with a high multivalent metal ion rejection rate through graphene oxide membranes. *J. Mater. Chem. A* **9**, 10672-10677 (2021).

26 Nair, R. R., Wu, H. A., Jayaram, P. N., Grigorieva, I. V. & Geim, A. K. Unimpeded permeation of water through helium-leak-tight graphene-based membranes. *Science* **335**, 442-444 (2012).

27 Huang, H. B. *et al.* Ultrafast viscous water flow through nanostrand-channelled graphene oxide membranes. *Nat. Commun.* **4**, 2979 (2013).

28 Han, Y., Xu, Z. & Gao, C. Ultrathin graphene nanofiltration membrane for water purification. *Adv. Funct. Mater.* **23**, 3693-3700 (2013).

29 Aher, A., Cai, Y. G., Majumder, M. & Bhattacharyya, D. Synthesis of graphene oxide membranes and their behavior in water and isopropanol. *Carbon* **116**, 145-153 (2017).

30 Liu, G., Jin, W. & Xu, N. Graphene-based membranes. *Chem. Soc. Rev.* **44**, 5016-5030 (2015).

31 Yang, E. *et al.* Tunable semi-permeability of graphene-based membranes by adjusting reduction degree of laminar graphene oxide layer. *J. Membrane Sci.* **547**, 73-79 (2018).

32 Yuan, Y. Q. *et al.* Enhanced desalination performance of carboxyl functionalized graphene oxide nanofiltration membranes. *Desalination* **405**, 29-39 (2017).

33 Zhang, H. *et al.* Construction of MoS2 composite membranes on ceramic hollow fibers for efficient water desalination. *J. Membrane Sci.* **592**, 117369 (2019).

34 Meng, B. C. *et al.* Fabrication of surface-charged MXene membrane and its application for water desalination. *J. Membrane Sci.* **623**, 119076 (2021).

35 Xue, Q. & Zhang, K. S. MXene nanocomposite nanofiltration membrane for low carbon and long-lasting desalination. *J. Membrane Sci.* **640**, 119808 (2021).

**Acknowledgements**

This work was supported by the National Natural Science Foundation of China (12075201, 12074341 and 12104394), the Science and Technology Planning Project of Jiangsu Province (BK20201428), the Scientific Research and Developed Funds of Ningbo University (No. ZX2022000015) and Hefei Advanced Computing Center.

**Author information**

Junfan Liu and Zonglin Gu contributed equally to this work.

**Author Contributions**

Y.T. and L.C. conceived the ideas. L.C. and Y.T. designed the experiments. J.L., P.L.,

L.L., R.Y., J.C. and Z.W. performed the experiments and prepared the data graphs. Y.T., L.C., Z.G., M.D., J.J. and L.Z. designed and performed the simulations. Y.T., L.C., Z.G. and J.L. co-wrote the manuscript. All authours discussed the results and commented on the manuscript.

**Additional information**

Supplementary information is available in the online version of the paper. Reprints and permissions information is available online at www.nature.com/reprints. Publisher's note: Springer Nature remains neutral with regard to jurisdictional claims in published maps and institutional affiliations. Correspondence and requests for materials should be addressed to Y.T. (ystu@yzu.edu.cn), L.C. (liang_chen05@126.com) and Z.G. (guzonglin@yzu.edu.cn).

**Competing interests**

The authors declare no competing financial interests.